\documentstyle[aps,prb,twocolumn,graphicx,latexsym]{revtex}
\newcommand{\bvec}[1]{{\bf #1}}
\newcommand{\mcdot}{\! \cdot \!}
\newcommand{\qi} {_{\bvec{q}}}
\newcommand{\mqi}{_{\bvec{-q}}}

\newcommand{\Acorr}{\langle \bvec{A} \qi \bvec{A} \mqi \rangle}
\newcommand{\hcorr}{\langle \bvec{h} \qi \bvec{h} \mqi \rangle}
\newcommand{\mcorr}{\langle \bvec{m} \qi \bvec{m} \mqi \rangle}
\newcommand{\etaA}{\eta_{\bvec{A}}}
\newcommand{\etah}{\eta_{\bvec{h}}}

\newcommand{\meffA}{m_{\rm eff}^{\bvec{A}}}
\newcommand{\meffh}{m_{\rm eff}^{\bvec{h}}}

\newcommand{\D}[1]{D#1}

\newcommand{\<}{\;\!\!}
\newcommand{\Prime}[1]{#1'}

\title{Anomalous scaling dimensions and stable charged fixed-point
  of type-II superconductors}
\author{J. Hove and A. Sudb{\o}}
\address{Department of Physics\\
         Norwegian University of Science and Technology,
         N-7491 Trondheim, Norway \\}
\date{\today}

\begin{document}
\maketitle
\begin{abstract}
  The critical properties of a type-II superconductor model are
  investigated using a dual vortex representation. Computing the
  propagators of gauge field $\bvec{A}$ and dual gauge field
  $\bvec{h}$ in terms of a vortex correlation function, we obtain the
  values $\etaA=1$ and $\etah=1$ for their anomalous dimensions. This
  provides support for a dual description of the
  Ginzburg-Landau theory of type-II superconductors in the continuum
  limit, as well as for the existence of a stable charged fixed point
  of the theory, not in the $3DXY$ universality class.
\end{abstract}

\draft 
\vskip1.5pc
Determining the universality class of the phase-transition in a 
system of a charged scalar field coupled to a massless gauge field, 
such as a type-II superconductor, has been a long-standing problem 
\cite{Halperin:L74}.
Analytical and numerical efforts have recently focused on the use of a
{\it dual} description of the Ginzburg-Landau theory (GLT) of type-II
superconductors, pioneered by Kleinert \cite{kleinert_bok},
in investigating the character of a proposed novel {\it stable} 
fixed point of the theory for a charged superconducting
condensate, in which case the $3DXY$ fixed point of the neutral
superfluid is rendered unstable
\cite{herbut_and_tesanovic,tesanovic_and_kiet_u1,kiet_large,asle_lt22}.
The dual formulation has also been employed to
investigate the possibility of novel broken symmetries in the
vortex liquid phase of such systems in magnetic fields
\cite{tesanovic_and_kiet_u1,kiet_large}.

The GLT is defined by a complex matter field $\psi$ coupled to a
massless fluctuating gauge field $\bvec{A}$ with a Hamiltonian
\begin{eqnarray}
\lefteqn{H_{\psi,\bvec{A}} = m_{\psi}^2 \left| \psi \right|^2 +
\frac{u_{\psi}}{2} \left| \psi \right|^4 + \left| \left( \nabla - i2e
    \bvec{A} \right) \psi \right|^2 +} \hspace{4.5cm} \nonumber \\
&& \frac{1}{2} \left( \nabla \times \bvec{A} \right)^2.
\label{hgl}
\end{eqnarray}
Here, $e$ is the electron charge, and $H_{\psi,{\bf A}}$ is invariant
under the local gauge-transformation $\psi \to \psi \exp(i \theta)$,
${\bf A} \to {\bf A} + \nabla \theta/2 i e$.  The GLT sustains stable
topological objects in the form of vortex lines and vortex loops, the
latter are the critical fluctuations of the theory
\cite{tesanovic_and_kiet_u1,kiet_large}. These objects are highly
nonlocal in terms of $\psi$, but a dual formulation offers a local
field theory for them.  The continuum dual representation of the
topological excitations, (in $D=3$ only), consists of a complex matter
field $\phi$ coupled to a {\em massive} gauge field $\bvec{h}$
\cite{kleinert_bok}, with coupling constant given by the dual charge
$e_d$, and with dual Hamiltonian  
\begin{eqnarray}
\lefteqn{H_{\phi,\bvec{h}} = m_{\phi}^2 \left| \phi \right|^2 +
\frac{u_{\phi}}{2} \left| \phi \right|^4 + \left| \left( \nabla - i e_d
    \bvec{h} \right) \phi \right|^2 +} \hspace{0.5cm} \nonumber \\
&& \frac{1}{2} \left( \nabla \times \bvec{h} \right)^2 + \frac{1}{2}
\left( \nabla \times \bvec{A}\right)^2 + ie \left( \nabla \times
  \bvec{h} \right) \cdot \bvec{A}.
\label{hdgl}
\end{eqnarray}
The massiveness of $\bvec{h}$ reduces the symmetry to a global
$U(1)$-invariance.  For details on how to obtain this dual Hamiltonian,
we refer the reader to the thorough exposition of
this presented in the textbook of Kleinert \cite{kleinert_bok1}. For
$e \neq 0$ the original GLT in Eq.  \ref{hgl} has a local gauge
symmetry, the dual theory in Eq.  \ref{hdgl} has a global $U(1)$
symmetry. In the limit $e \to 0$, $\bvec{A}$ decouples from $\psi$ in
Eq.  \ref{hgl}, $H_{\psi}$ describes a {\em neutral superfluid}, and
the symmetry is reduced to global $U(1)$. The dual Hamiltonian
$H_{\phi,\bvec{h}}$ describes a charged superfluid coupled to a
massless gauge field $\bvec{h}$ with coupling constant $e_d$, and the
global symmetry is extended to a local gauge symmetry. Hence, when $e
\to 0$, {\it the dual of a neutral superfluid is isomorphic to a
superconductor}. Integrating out the ${\bf A}$ field in Eq. \ref{hdgl}
produces a mass-term $e^2 {\bf h}^2/2$, where an exact
renormalization-group equation for the mass of ${\bf h}$ is given by
$\partial e^2/\partial \ln l = e^2$ \cite{calanII}.  Therefore, when
$e \neq 0$, then $e ^2 \to \infty$ as $l \to \infty$.  This supresses
the dual gauge field, and the resulting dual theory is a pure
$|\phi|^4$-theory. Hence, in the long-wavelength limit, {\it the dual of a
superconductor is isomorphic to a neutral superfluid} \cite{kleinert_bok}.

In this paper, we obtain the anomalous scaling dimensions $\etaA$ of
the gauge field \cite{herbut_and_tesanovic,olsson_prl_80_1998}, as
well as $\etah$ of the dual gauge field, not previously considered,
directly from large-scale Monte-Carlo simulations. At a $3DXY$
critical point,$\etaA = \etah=0$. We find that $(\etaA = 1, \etah=0)$
when $e \neq 0$, and that $(\etaA=0, \etah=1)$, when $e=0$.  We also
contrast the anomalous dimension of the dual mass field $\phi$ at the
dual charged (original neutral) and dual neutral (original charged)
fixed points, obtaining $\eta_{\phi}=-0.24$ in the former case, and
$\eta_{\phi}=0.04$ in the latter.

A duality transformation, to a set of interacting vortex loops, is
performed on the London/Villain approximation to the GLT. In this
approximation the partition function is
\begin{eqnarray}
\lefteqn{Z(\beta,e) = \int \D{\bvec{A}} \D{\theta} \sum_{\{\bvec{n}\}} 
\exp\left[ -\sum_{\bvec{x}} \left\{ \frac{1}{2} 
\left(\Delta \times \bvec{A} \right)^2 
+ \right. \right.} 
\hspace{2.25cm} \nonumber \\
&&\left.\left.\frac{\beta}{2}\left(\Delta \theta - e\bvec{A} - 2\pi \bvec{n}
    \right)^2 \right\} \right].
\label{orgtheory}
\end{eqnarray}
Here, $\theta$ is the local phase of the superconducting order
parameter $\psi$, while $\bvec{n}$ is an integer-valued {\em
  velocity field} (not vortex field) introduced to make the Villain
potential $2 \pi$-periodic. The symbol $\Delta$ denotes a lattice
derivative. Amplitude fluctuations are neglected in this approach. The
validity of this approximation for $3D$ systems, has recently been
investigated in detail, both numerically and analytically
\cite{phase_ref}.

An auxiliary velocity field $\bvec{v}$ linearises the kinetic energy.
Performing the $\theta$-integration constrains $\bvec{v}$ to satisfy
the condition $\Delta \cdot \bvec{v} = 0$, explicitly solved by
writing $\bvec{v} = \Delta \times \bvec{h}$, where $\bvec{h}$ is
forced to integer values by the summation over $\bvec{n}$. Introducing
an integer-valued {\em vortex field} $\bvec{m} = \Delta \times
\bvec{n}$, and using Poisson's summation formula, we find

\begin{eqnarray}
    S(\bvec{A},\bvec{h},\bvec{m}) &=& 
    \sum_{\bvec{x}} \bigg\{
      2\pi i \bvec{m}\cdot\bvec{h} + \frac{1}{2\beta} \left( \Delta \times
        \bvec{h}\right)^2 \nonumber \\
&& + ie \left(\Delta \times \bvec{h}\right) \bvec{A}
 + \frac{1}{2}\left(\Delta \times \bvec{A} \right)^2 \bigg\} .
\label{ahm-action}
\end{eqnarray}
Integrating the gauge field in Eq. \ref{ahm-action} produces a mass
term $e^2\bvec{h}^2/2$, giving an effective theory containing the
vortex field $\bvec{m}$ coupled to a {\em massive} gauge field
$\bvec{h}$
\begin{eqnarray}
  \lefteqn{
    Z(\beta,e) = \int\D{\bvec{h}} \sum_{\{\bvec{m}\}} 
 \prod_{\bvec{x}} \delta_{\Delta \cdot \bvec{m},0}
  \exp \bigg[
  -\sum_{\bvec{x}}\bigg\{ 2\pi i \bvec{m} \mcdot \bvec{h}\,+} 
\hspace{2.25cm} \nonumber \\ 
&&
\frac{e^2}{2}
    \bvec{h}^2 + \frac{1}{2\beta} \left( \Delta \times \bvec{h}
    \right)^2 \bigg\} \bigg].
\label{dual}
\end{eqnarray}
The variables $\bvec{m}$ in Eq. \ref{dual} describe a set of
interacting vortices, where the interactions are mediated through the
gauge field $\bvec{h}$. The variables in Eq. \ref{dual} are defined 
on a lattice which is dual to the lattice from Eq. \ref{orgtheory}, 
and the behavior with respect to temperature is inverted in the new 
variables. The $\theta$ field in Eq. \ref{orgtheory} describes 
{\em order}, while the $\bvec{m}$ field represents the topological 
 excitations of the $\theta$ field. These excitations destroy 
superconducting coherence, and hence quantify 
{\em disorder} \cite{kleinert_bok1}. 

Integrating out the $\bvec{h}$ field in Eq. \ref{dual}, we obtain
the Hamiltonian employed in the present simulations,
\begin{eqnarray}
H(\bvec{m}) &=& - 2\pi^2 J_0 \sum_{\bvec{x_1},\bvec{x_2}}
\bvec{m}(\bvec{x_1}) V(\bvec{x_1} - \bvec{x_2}) \bvec{m}(\bvec{x_2}), 
\label{finalH} \\
V(\bvec{x}) &=& \sum_{\bvec{q}} \frac{e^{-i\bvec{q} \cdot \bvec{x}}}{4
  \sum_{\mu} \sin^2 \left( \frac{q_{\mu}}{2} \right) + \lambda^{-2}} \
\label{finalV}.
\end{eqnarray}
In Eq. \ref{finalV}, the charge $e$ and lattice-spacing $a$ have 
both been set to unity, and $\lambda$ is the bare London penetration 
depth. At every MC step, we attempt to insert a loop of unit vorticity 
and random orientation. A new energy is calculated from Eq. 6, and the 
proposed move is accepted or rejected according to the Metropolis 
algorithm. This procedure ensures that the vortex lines of the system 
always form closed loops of random size and shape \cite{kiet_large}.  
In all simulations, a system size of $40 \times 40 \times 40$ was 
used, and up to $1.5 \cdot 10^5$ sweeps over the lattice per temperature 
were used.

To investigate the properties of $\bvec{A}$ and $\bvec{h}$ at the
charged critical point of the original theory, Eq. \ref{hgl}, we have
calculated the correlation functions $\Acorr$ and $\hcorr$ in terms of
vortex correlations, obtaining
\begin{eqnarray}
\Acorr &=& \frac{1}{\left| \bvec{Q} \right|^2 + m_0^2}
\left(1 + \frac{4\pi^2 \beta m_0^2
    G(\bvec{q})}{\left|\bvec{Q}\right|^2 \left( 
  \left| \bvec{Q} \right|^2 + m_0^2\right)}  \right), \label{acorr}\\
\hcorr &=& \frac{2\beta}{\left| \bvec{Q} \right|^2 + m_0^2} \left( 1 -
  \frac{2\beta \pi^2 G(\bvec{q})}{\left| \bvec{Q} \right|^2 + m_0^2}
  \right), \label{hcorr} 
\end{eqnarray}
where $G(\bvec{q}) = \mcorr$, $m_0 = \lambda^{-1}$ and $Q_{\mu} = 1 -
e^{-i\bvec{q} \cdot \hat{\mu}}$. All correlation functions have been
calculated in the {\em transverse gauge} $\nabla \cdot \bvec{A} =
\nabla \cdot \bvec{h} = 0$. Both of the fields  ${\bf h}$ and ${\bf A}$ 
are renormalized by vortex fluctuations, albeit in quite
different ways.

Invoking the standard form $\left(q^2 + m_{\rm eff}^2\right)^{-1}$ for
the correlation functions in the immediate vicinity of the critical
point in the limit $q \to 0$, we find the following expressions for
the effective masses,
\begin{eqnarray}
\left(\meffA\right)^2 &=& \lim_{q\to 0}  \frac{m_0^2}{1 + 4 \pi^2 \beta
      G(\bvec{q})q^{-2}}, \label{meffa}\\
\left(\meffh\right)^2 &=& \lim_{q\to 0}  \frac{m_0^2}{2\beta\left(1 -
      \frac{2 
      \pi^2 \beta G(\bvec{q})}{m_0^2}\right)}. \label{meffh}
\end{eqnarray}
When {\em $e \neq 0$} the correlation function for $\bvec{A}$ assumes
the form
\begin{equation} 
\Acorr \propto \frac{1}{q^{2 - \etaA}} \label{etaa}
\end{equation}
at the critical point. To determine $\eta_{{\bf A}}$, we compute the 
vortex correlator $G(q)$. For $\lambda << L = 40$ , we expect the 
following behaviour for $G(\bvec{q})$ in the limit $q \to 0$,
\begin{eqnarray}
T < T_c &\Rightarrow & G(\bvec{q}) \propto q^2, \label{glimita}\\ 
T = T_c &\Rightarrow & G(\bvec{q}) \propto \ q^{\eta}, \label{glimitb}\\
T > T_c &\Rightarrow & G(\bvec{q}) \propto \ C(T). \label{glimitc}
\end{eqnarray}
When these limiting forms are inserted in Eq. \ref{meffa}, we see that
for $T \le T_c$, $\meffA$ will be finite through the Higgs Mechanism
(Meissner effect).  For $T \geq T_c$ we will have $\meffA = 0$ as in
the normal case of a massless photon. Assuming $G(q) \propto q^{\eta}$
precisely at the critical point, it is seen that $\eta$ corresponds to
$\etaA$ from Eq.  \ref{etaa}. {\it We thus identify the scaling power
  of $G(\bvec{q})$ at the critical point with the anomalous dimension
  of the massless gauge field ${\bf A}$.}

\begin{figure}[htbp]
\hfill\scalebox{0.34}{\rotatebox{-90.0}{\includegraphics{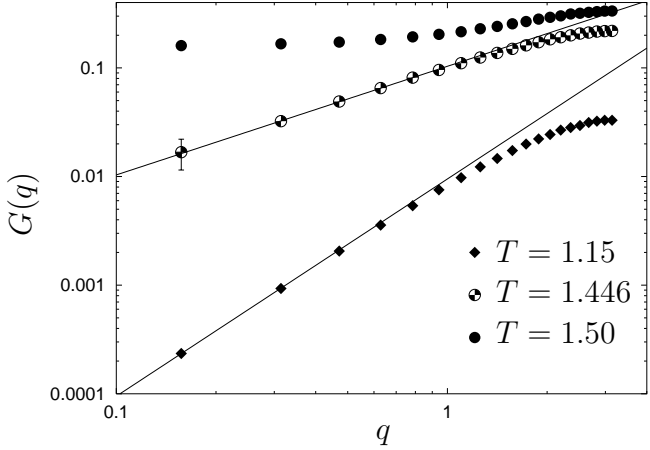}}}
\newline
\setlength{\unitlength}{1cm}
\begin{picture}(0,0)
\put(4.9,0.35)  {\large $q$}
\put(-0.0,3.05) {\large \rotatebox{90}{$G(q)$}}
\put(6.5,2.70)  {\large $T = 1.15$}
\put(6.5,2.165) {\large $T = 1.446$}
\put(6.5,1.62)  {\large $T = 1.50$}
\end{picture}
\caption{\label{Gfig} log-log plot of $G(q)$ for the three alternatives
  in Eq. \ref{glimita} - \ref{glimitc}, with $\lambda = a/2$. For this
$\lambda$, $T_c=1.446$. Apart from the point $q = q_{\min}, T=1.446$
the error bars are smaller than the symbols used.}
\end{figure}

\begin{figure}[htbp]
\hfill\scalebox{0.34}{\rotatebox{-90.0}{\includegraphics{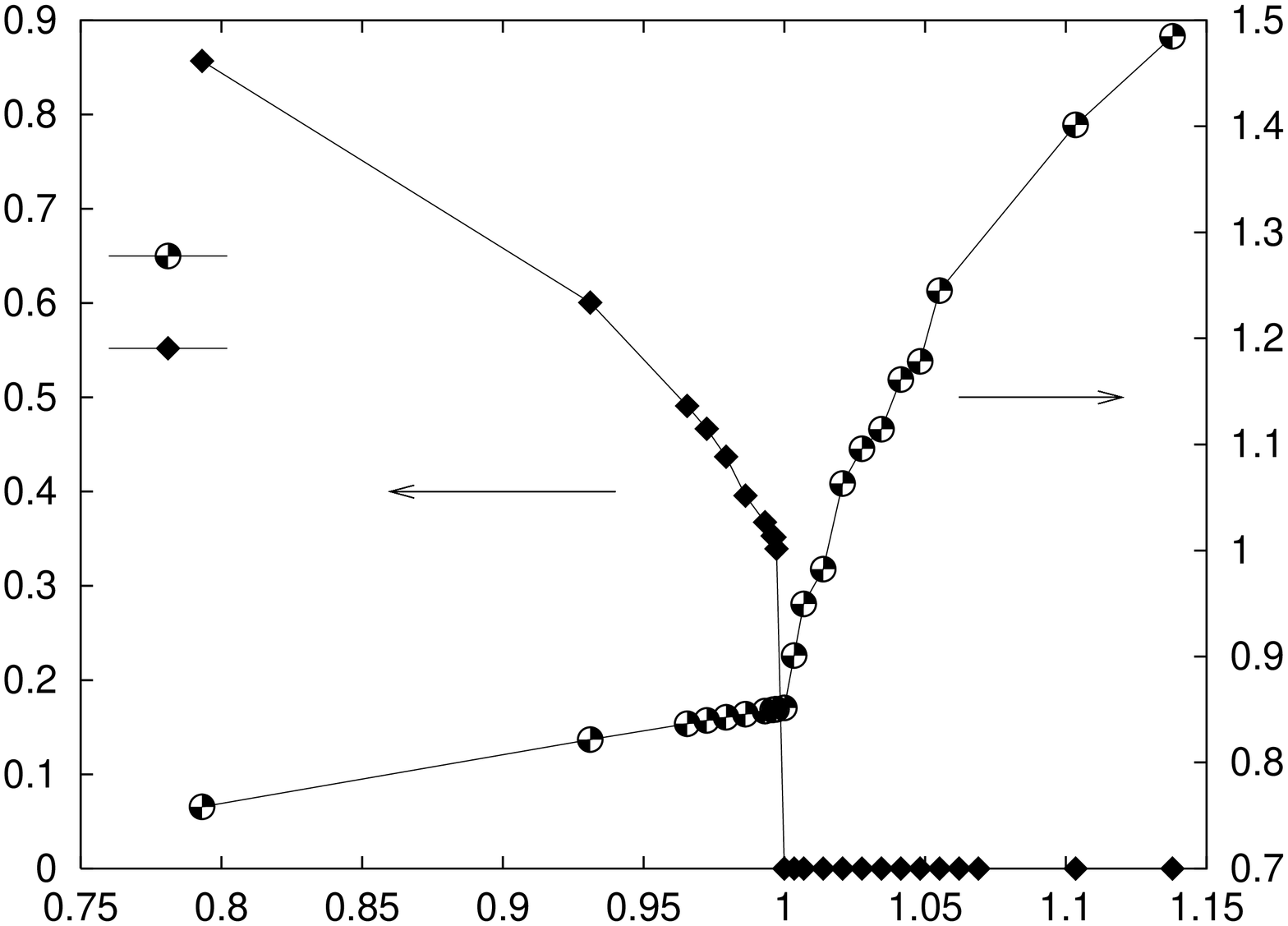}}}
\newline
\setlength{\unitlength}{1cm}
\begin{picture}(0,0.5)
\put(4.05,0.50){\large $T/T_c$}
\put(-0.1,3.7){\rotatebox{90}{\Large $m_{\bvec{A}}/m_0$}}
\put(8.5 ,3.7){\rotatebox{90}{\Large $m_{\bvec{h}}/m_0$}}
\end{picture}
\caption{\label{mhfig} $\meffA/m_0$ and $\meffh/m_0$ as functions of $T$.}
\end{figure}

All three limiting forms Eqs. \ref{glimita}-\ref{glimitc} are shown in
Fig. \ref{Gfig}. The gauge field masses $\meffh$ and $\meffA$ in Eqs.
\ref{meffa} and \ref{meffh}, are shown in Fig. \ref{mhfig}. At the
critical point $G(q) \propto q$, so that $\etaA = 1$. Note that, while
$\meffA$ vanishes at $T=T_c$, $\meffh$ is finite but non-analytic.
{\em As a result of the vortex loop blowout, the screening properties
  of the vortices are dramatically increased, and $\meffh$ increases
  sharply.}

To find $\etah$ independently, we consider first the uncharged case
$\lambda \to \infty$, $m_0 \to 0$.  First, at an intermediate step 
in the transformation Eqs. \ref{orgtheory} - \ref{dual}, the action reads
\begin{equation}
S(\beta,e) = -\sum_{\bvec{x}} \left\{ \frac{1}{2\beta}\bvec{l}^2 +
  ei\bvec{A} \cdot 
  \bvec{l} + \frac{1}{2} \left(\nabla \times \bvec{A}\right)^2
\right\}.
\label{lfield}
\end{equation}
Here, $\bvec{l}$ is an integer field of closed current loops. Setting
$e=0$ in Eq. \ref{dual}, the action of the dual Villain model is
obtained,
\begin{equation}
\tilde{S}_V(\beta,\Gamma) \<\< =
  \<\<-\<\<\sum_{\bvec{x}}\<\<\left\{\<2\pi i \bvec{m} 
  \<\<\cdot\<\< \bvec{h} \<+\< \frac{1}{2\beta} \left( \Delta
  \<\times\< \bvec{h} 
  \right)^2 \<+\<
 \frac{\Gamma}{2}\bvec{m}^2\<\<\right\}\<\<\<\<.
\label{kfield}
\end{equation}
Here, a term $\Gamma\bvec{m}^2/2$ has been added, and
$\tilde{S}_V(\beta,\Gamma)$ corresponds to the Villain-action in the
limit $\Gamma \to 0$. However, it is physically reasonable to propose
that the limit $\Gamma \to 0$ is non-singular, since the added term is
short-ranged. It should therefore be an irrelevant perturbation, in
renormalization group sense, to the long-ranged Biot-Savart 
interaction governing 
the fixed point, which is  mediated by $\bvec{h}$. 
Rescaling $\bvec{h} \to \bvec{h}e /2\pi$ in Eq.
\ref{kfield}, we have \cite{dasgupta_prl_47_1981}$Z(\beta,e) =
\tilde{Z}_V\left(e^2/4\pi^2,1/2\beta)\right)$, leaving Eqs.
\ref{lfield} and \ref{kfield} interchangeable; $\etah$ from Eq.
\ref{kfield} should have the same value as $\etaA$ from Eq.
\ref{lfield}.  The above is demonstrated by our simulations
based on Eqs. \ref{finalH}-\ref{hcorr}, {\it which are independent 
of the proposed form Eq. \ref{kfield}}. 

\begin{figure}[htbp]
\hfill\scalebox{0.34}{\rotatebox{270.0}{\includegraphics{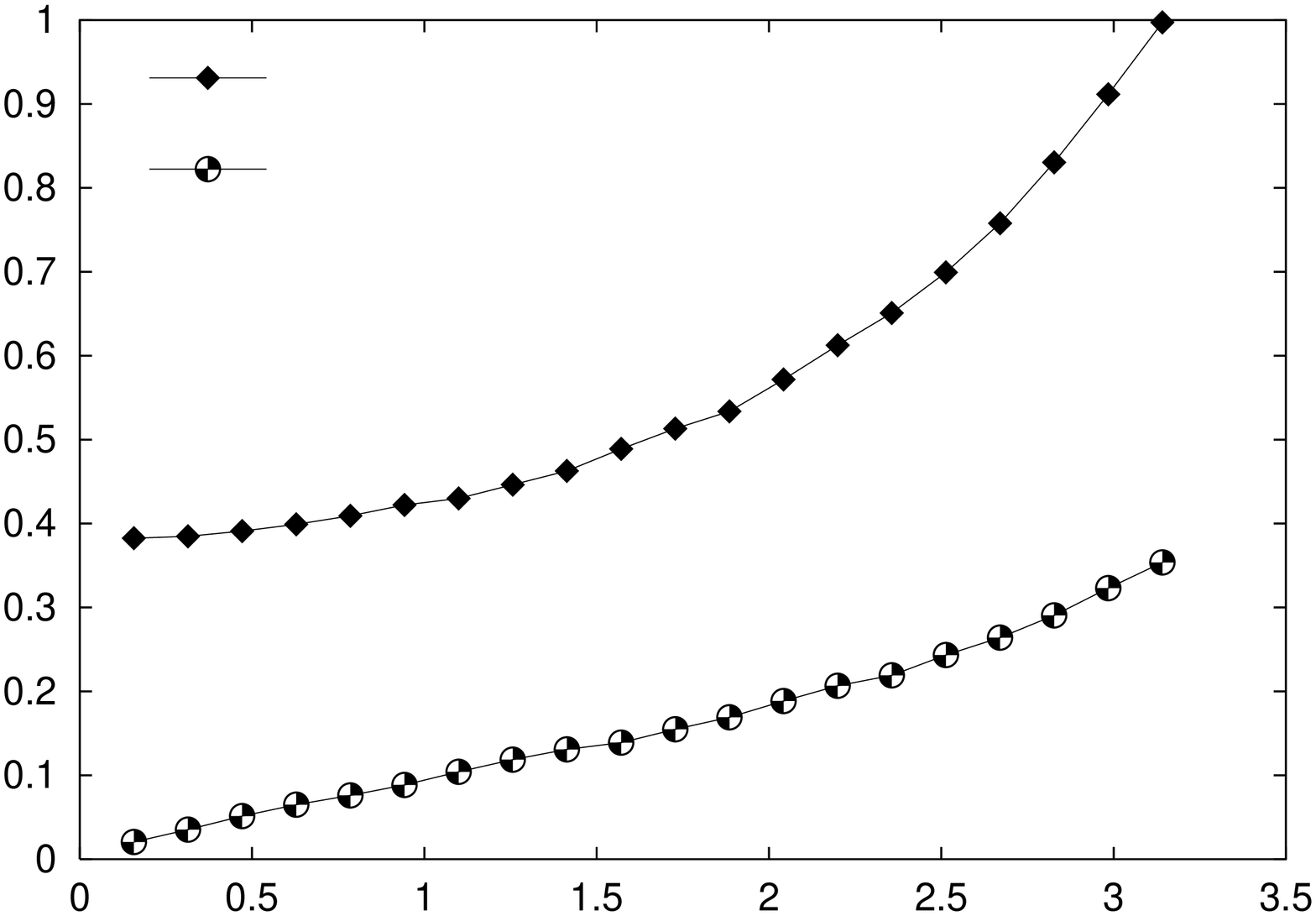}}}
\newline
\setlength{\unitlength}{1cm}
\begin{picture}(0,0)
\put(4.8,0.4) {\large$q$}
\put(-0.0,2.6){\rotatebox{90}{\large $q^2\hcorr$}}
\put(2.3,5.65){\large $T=2.75$}
\put(2.3,5.12){\large $T=3.00$}
\end{picture}
\caption{\label{Hfig} $q^2\langle \bvec{h} \qi
  \bvec{h} \mqi \rangle$ for two different $T$.  For $\lambda =
  \infty$, $T_c=3.00$.}
\end{figure}

To determine $\etah$ we study the correlation function $\hcorr$ (Eq.
\ref{hcorr}) in the limit $m_0 \to 0$. At the uncharged fixed point of
the original theory, which is the charged fixed point of the dual
theory, we have $\lim_{q \to 0} 2 \pi \beta^2 G({\bf q}) = (1-C_2(T))q^2+..., 
q^2-C_3(T) q^{2 +\eta_{\bf h}}+...$, and $q^2-C_4(T) q^4+...$, for $T<T_c, T=T_c$, 
and $T \geq T_c$, respectively. Here, $C_2(T)$ corresponds to the helicity 
modulus (superfluid density) \cite{kiet_llm}, $C_3(T)$ is a critical amplitude, 
and $C_4(T)$ is the inverse of the mass of the dual gauge field for 
$T \geq T_c$. Correspondingly, we have $\lim_{q \to 0} <{\bf h_{\bf q} h_{-\bf q}}> =
2 \beta C_2/q^2, ~~ 2 \beta C_3/q^{2-\eta_{\bf h}}$, and $2 \beta C_4$, for
$T<T_c, T=T_c$, and $T \geq T_c$, respectively. Note that ${\bf h}$ is massless 
for $T< T_c$, while it is massive for $T>T_c$, the dual system exhibits a ``dual 
Meissner-effect'' for $T \geq T_c$.  At $T=T_c$, we have 
$q^2 ~~ \hcorr \simeq C_3(T) q^{\eta{\bvec{h}}}$. A plot of 
$q^2 ~~ \hcorr$ is 
shown in Fig. \ref{Hfig}. A linear behaviour at $T = T_c$ is found, implying 
that $\etah = 1$ when $e=0$. Since $\etah = 1$ in the uncharged case, this 
provides further support for the Hamiltonian Eq. \ref{hdgl}.

We now set $e \neq 0$. The gauge field $\bvec{h}$ becomes massive
via the term $e^2\bvec{h}^2/2$, which appears after integrating out
the $\bvec{A}$ field in Eq. \ref{hdgl}. In this case, $\lim_{q \to 0}
\hcorr = 2\beta / m_0^2$ from Eq. \ref{hcorr}, and $\bvec{h}(r)$
would naively have the trivial scaling dimension $\left( 2 - d \right)
/ 2$. However, the mass term offers us a freedom in assigning
dimensions to $e$ and $\bvec{h}$, by introducing renormalization
$Z$-factors, here $\Prime{e} = Z_{\bvec{h}}^{1/2}e$ and
$\Prime{\bvec{h}} = Z_{\bvec{h}}^{-1/2}\bvec{h}$.

Prior to integrating out $\bvec{A}$ in Eq. \ref{hdgl}, the mass 
appears in the term 
$ie \left( \nabla \times \bvec{h} \right) \cdot \bvec{A}$.  
Integration of the $\phi$ field, partial or complete, can only 
produce $(\nabla/i - e_d {\bf h})$-terms. In particular, 
this must hold during  integration of fast 
Fourier-modes of the $\phi$ field. Thus, the term 
$ i ( {\bf \nabla} \times {\bf h}) \cdot {\bf A}$
{\em is renormalisation group invariant}, i.e. its prefactor must be
dimensionless. In terms of scaled fields, at the charged fixed point
of the original theory, we have $\Prime{\bvec{A}} =
Z_{\bvec{A}}^{-1/2}\bvec{A}$, with $Z_{\bvec{A}} \propto l^{\etaA}$,
$\etaA = 1$\cite{calanII}. For $\bvec{h}$, we use $Z_{\bvec{h}}
\propto l^{\Delta}$, where $\Delta$ is {\em not} an anomalous scaling
dimension ($\bvec{h}$ is massive, cf. Fig.  \ref{mhfig}), but rather a
contribution to the engineering dimension of $\bvec{h}$. Inserting
this into the crossterm $ie \left( \nabla \times \bvec{h} \right)
\cdot \bvec{A}$, we find the scaling dimension $\left( \etaA + \Delta
\right)/2 - 1$, which must vanish. This gives the constraint
$\Delta = 1$ to avoid conflicting results for $\eta_{\bvec{A}}$. 

Remarkably, therefore, the scaling dimension of $\bvec{h}$ at $T=T_c$ is 
the same in both cases $m_0=0$ and $m_0 \neq 0$. The results for $\etaA$ 
and $\etah$ in the previous paragraphs, are summed up in Table \ref{tabell}. 

We next consider the distribution of vortex loop sizes in
the model Eq.  \ref{finalV}, connecting the
vortex loop distribution to the anomalous dimension of $\phi$ at
$T_c$ {\it both for the case $e=0$ and $e \neq 0$}. During the
simulations, we sample the distribution of loop-sizes $D(p)$, where
$p$ is the perimeter of a loop. This distribution function can be 
fitted to the form \cite{kiet_large,tesanovic_and_kiet_u1}
\begin{equation}
 D(p) \propto p^{-\alpha} e^{-\beta p \varepsilon(T)},
\label{formuladp}
\end{equation} 
where $\varepsilon(T)$ is an effective line-tension for the loops.
Figures showing the qualitative features of $D(p)$ can be found in
Ref. \onlinecite{kiet_large}. The critical point is characterised by a
vanishing line-tension, and close to the critical point we find that
$\varepsilon(T)$ vanishes as $\varepsilon(T) \propto \left| T - T_c
\right|^{\gamma_{\phi}}$.

The vortex loops are the topological excitations of the GL and 3DXY
models, at the same time they are the real-space representation of the
Feynman diagrams of the dual field theory. By sampling $D(p)$, we 
obtain information about
the dual field $\phi$, particularly $\gamma_{\phi}$ can be identified
as a {\em susceptibility} exponent for the $\phi$
field\cite{kiet_large}. Using the scaling relation $\gamma_{\phi} =
\nu_{\phi} \left( 2 - \eta_{\phi}\right)$, and the important
observation that even at the charged dual fixed point $\nu_{\phi} =
\nu_{3DXY}$ \cite{kiet_large}, this also gives us a value for the
anomalous scaling dimension $\eta_{\phi}$ when we use the value
$\nu_{3DXY}=0.673$ \cite{hasenbusch_99}.

In Ref. \onlinecite{kiet_large} the vortex loops of the 3DXY model
have been studied meticulously, yielding the value $\eta_{\phi}(0) =
-0.18 \pm 0.07$. Since the dual of this model is isomorphic
to a superconductor, $\eta_{\phi}(0)$ should be similar to
$\eta_{\psi}(e)$ of the original GLT.

\begin{figure}[htbp]
\hfill\scalebox{0.34}{\rotatebox{270.0}{\includegraphics{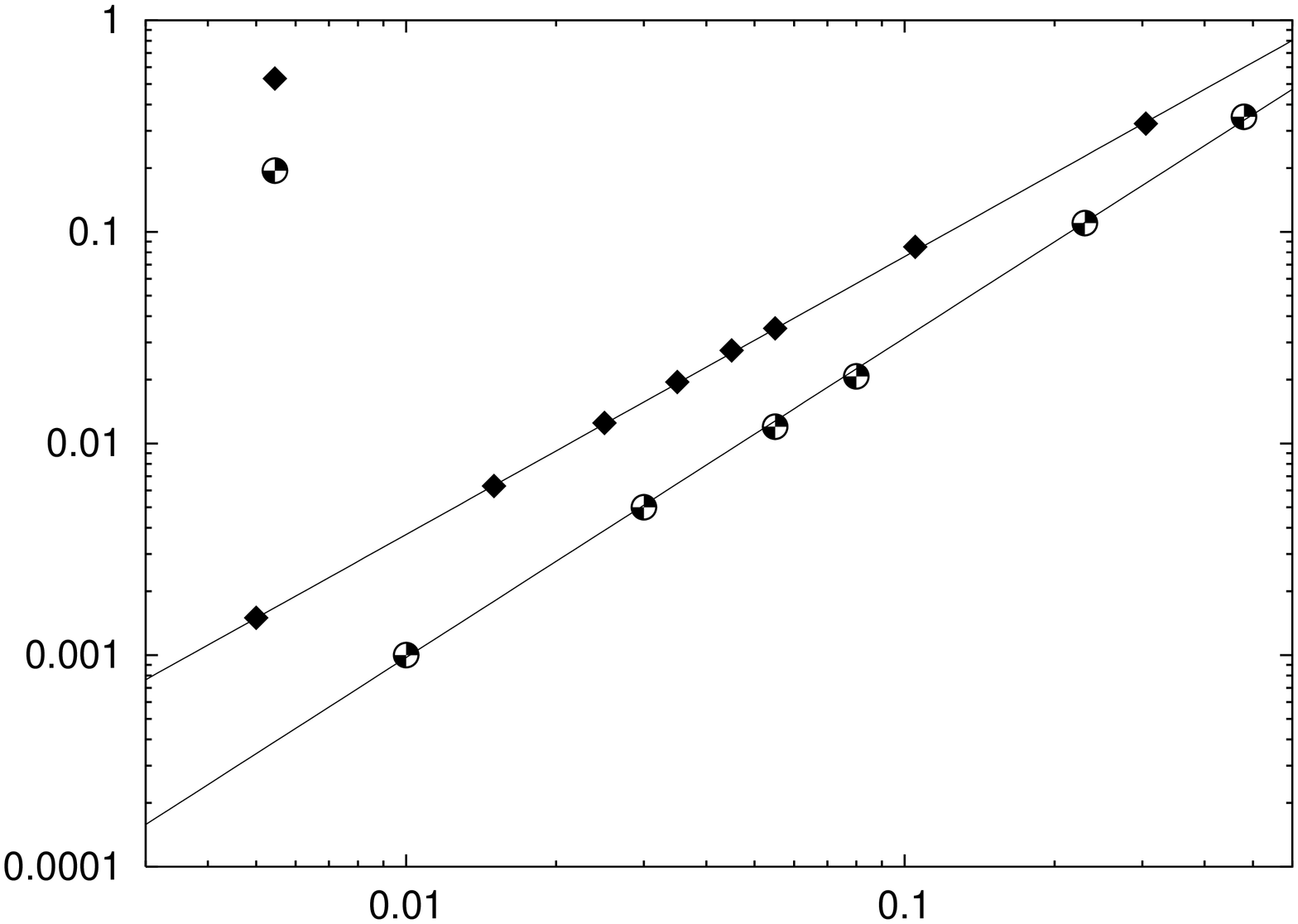}}}
\newline
\setlength{\unitlength}{1cm}
\begin{picture}(0,0)
\put(3.7,0.3){\large$\log \left| T - T_c \right|$}
\put(0.35,2.8){\rotatebox{90}{\large $\log \varepsilon(T)$}}
\put(2.5 , 5.65){\large $\lambda L^{-1} = 0.0125$}
\put(2.5 , 5.06){\large $\lambda L^{-1} = \infty$}
\end{picture}
\caption{\label{ltfig}
   $\ln \varepsilon(T)$ as a function of $\ln \left| T - T_c \right|$.
  The upper line shows the charged case with finite $e$, and the lower
  line shows the neutral case with $e = 0$. The slopes of the two
  straight lines are $\gamma_{\phi} = 1.315$ and $\gamma_{\phi} =
  1.51$, corresponding to the anomalous dimensions $\eta_{\phi} =
  -0.24$ (neutral, i.e. dual charged) and $\eta_{\phi} = 0.04$ 
  (charged, i.e. dual neutral), respectively.}
\end{figure}

We have studied the vortex loop distribution in both the neutral and
the charged case. In the former case we find $\eta_{\phi} \simeq
-0.24$, in good agreement with Ref. \onlinecite{kiet_large}. In the
latter case the dual theory has a $U(1)$ symmetry, and we would expect
to find $\eta_{\phi} = \eta_{3DXY}$. The exponent $\eta_{3DXY}$ has
recently been determined with great accuracy to $\eta_{3DXY} =
0.038$\cite{hasenbusch_99}, whereas we find $\eta_{\phi} \simeq 0.04$
which compares well with this value.  Fig. \ref{ltfig} shows
$\varepsilon(T)$ for both the charged and uncharged models. {\it It is
  evident that they belong to two different universality classes.}

In the case $e \neq 0$, {\it which corresponds to the dual neutral
  case}, the inverse $\phi$-propagator is given by $G^{-1} = q^2 +
\Sigma(q)$, where $\Sigma$ is a self-energy, and $\Sigma(q) \sim
q^{2-\eta}$ by definition. This gives a leading order behavior $G \sim
1/q^{2-\eta}$ provided $\eta > 0$, and we find $\eta = 0.04$ for this
case.  On the other hand, for the case $e=0$, {\it which corresponds
  to the dual charged case}, dual gauge field fluctuations alter the
physics, softening the long-wavelength $\phi$ field fluctuations. We
obtain $G^{-1} = q^4 +\Sigma(q)$, again with $\Sigma(q) \sim
q^{2-\eta}$, which now gives a leading order behavior $G \sim 1/q^{2 -
  \eta}$, provided $\eta >-2$. Our result $\eta =-0.24$ for the case
$e=0$ (dual charged) is consistent with this, and also with the
absolute bounds $\eta > 2-D = -1$, in $D=3$.

A consequence of the above is that in $D=3$ dimensions, $\lambda \sim
\xi^{(D-2)/(2-\etaA)}=\xi$ at the charged critical point, in contrast
to $\lambda \sim \sqrt{\xi}$ at the $3DXY$ neutral critical point.
Since our results have been obtained directly by MC simulations, they
are valid beyond all orders in perturbation theory.

This work has been supported by a grant of computing time from 
Tungregneprogrammet, Norges Forskningsr{\aa}d.  We thank Para//ab for valuable 
assistance in optimizing our computer codes for use on the Cray Origin 
2000, and Zlatko Te\v{s}anovi\'{c} for many useful discussions.

\begin{table}[htbp]
\begin{tabular}{l|ll|ll}
$m_0$  & $\etaA$ & FP, original theory & $\etah$ & FP, dual theory  \\ \hline 
0      &  0      & Neutral 3DXY              &    1    & Charged \\
Finite &  1      & Charged                   &    0    & Neutral 3DXY 
\end{tabular}
\caption{\label{tabell} Values of $\etaA$ and $\etah$ at the stable 
neutral and charged critical points of the original and dual theories.
FP is an abbreviation for fixed point.}
\end{table}
\end{document}